\documentclass[final,3p,times]{elsarticle}

\usepackage{amssymb}
\usepackage{amsmath}

\usepackage{lineno} 
\usepackage{caption}
\usepackage{wrapfig}

\usepackage{hyperref}
\hypersetup{
colorlinks=true,
linkcolor=blue, 
urlcolor=black,
citecolor=red
}

\journal{Journal of Subatomic Particles and Cosmology}

\begin{document}

\begin{frontmatter}



\title{Anisotropic Flow in Ultra-Central Pb\textendash Pb Collisions at $\sqrt{\mathrm{s_{NN}}}=5.36$ TeV with ALICE}

\author[1]{Iris Likmeta}
\author{on behalf of the ALICE Collaboration}
\affiliation[1]{organization={University of Houston},
             addressline={3507 Cullen Blvd},
             city={Houston},
             postcode={77204},
             state={TX},
             country={USA}}

\begin{abstract}
Anisotropic flow measurements in heavy-ion collisions are sensitive to the spatial distribution of the initial state, and quark--gluon plasma transport properties such as the shear viscosity to entropy density ratio $(\eta/s)$. State-of-the-art relativistic hydrodynamic models successfully describe such flow measurements over a wide centrality range. However, the hydrodynamic description of anisotropic flow deviates from the experimental data in ultra-central collisions (UCC), where the average geometric anisotropy of the system becomes small and fluctuations dominate the initial-state geometry. This discrepancy constitutes the UCC puzzle, as the expected hierarchy of flow harmonics is not fully reproduced by current modeling approaches. Probing towards ultra-central collisions, effects on flow fluctuations due to the initial spatial anisotropies are suppressed. The measured flow can be explained by quantum fluctuations on the energy distribution of {$^{208}$Pb} nuclei. An octupole deformation of the {$^{208}$Pb} nuclei has been proposed as a remedy to improve the modeling of the measured $v_{3} \{2\} / v_{2} \{2\}$ ratio. In this contribution, we present measurements of $v_{3}\{2\}/v_{2}\{2\}$ ratio in ultra-central Pb\textendash Pb collisions at $\sqrt{s_{\mathrm{NN}}} = 5.36$ TeV with ALICE Run 3 detector and compare them with recent hydrodynamic model calculations.
\end{abstract}




\end{frontmatter}



\section{Introduction}
\label{sec1}

While the quark--gluon plasma exists only fleetingly, its properties can be inferred from the distribution, correlations, and interactions of the particles that emerge as it cools and hadronizes. Among these properties, collective behavior \cite{Ollitrault1992, Heinz2013FlowReview} stands out as a definitive signature of the strongly coupled nature of the medium. In QGP studies, collectivity is primarily investigated through measurements of azimuthal anisotropy. These anisotropies appear in the distribution of particle momenta with respect to the reaction plane. Precisely, the initial spatial asymmetry of the overlapping nuclei in non-central collisions generates pressure gradients that convert the initial geometric anisotropy into momentum-space anisotropy during the system’s evolution. Anisotropic flow is quantified through the coefficients of the Fourier expansion of the azimuthal momentum distribution of produced particle' momenta, as expressed in Eq.~(\ref{eq:dN}):   
\begin{equation} \label{eq:dN} 
    \frac{{\rm d}N}{{\rm d}\varphi} \propto 1 + 2\sum_{n=1}^{\infty} v_{n} \cos [n(\varphi-\Psi_n)]
\end{equation}
Here, $\varphi$ denotes the azimuthal angle of the emitted particle, and $\Psi_n$ is the $n$th-order symmetry plane angle, which reflects the orientation of the initial spatial anisotropy of the collision. The Fourier coefficients, $v_n=\left\langle \cos\left[n(\varphi-\Psi_n)\right]\right\rangle$, quantify the magnitude of the anisotropic flow for the $n$th harmonic ~\cite{Voloshin:1994mz}, where $\langle\cdots\rangle$ denotes an average over particles and events. The first-, second-, and third-order coefficients, $v_1$, $v_2$, and $v_3$, are referred to as the \textit{directed}, \textit{elliptic}, and \textit{triangular} flow, respectively.

Previous measurements \cite{ALICE:Adam_2016} of Fourier flow coefficients  exhibit good agreement with hydrodynamic model predictions all across the centrality range. However, probing towards the most central collisions, the picture seems to be completely different. Ultra-central collisions refer to most head-on heavy-ion collisions, typically in the top 1\% of centrality, where the overlap between the two nuclei is almost perfectly spherical. In such collisions, the average geometric overlap is nearly symmetric, so one would naively expect the elliptic flow coefficient, which arises from the almond-shaped overlap in non-central collisions, to be very small (ideally zero). Triangular flow, on the other hand, originates mostly from event-by-event fluctuations in the positions of nucleons within the nuclei, rather than from the average geometry.

The puzzle arises from experimental observations in ultra-central Pb\textendash Pb collisions when the ratio $v_3\{2\} / v_2\{2\}$ approaches or even exceeds unity. This means that triangular flow becomes as large as, or larger than, elliptic flow in collisions where the average geometry should be nearly spherical. Such a result is counterintuitive because one would expect $v_2\{2\}$ to dominate in non-central collisions, and $v_3\{2\}$ to remain relatively small, even when including fluctuations. Recent work in~\cite{Giannini_2023} highlights a significant tension between experimental measurements and theoretical predictions in ultra-central heavy-ion collisions. In particular, the measured ratio of triangular to elliptic flow, $v_3\{2\}/v_2\{2\}$, approaches or even exceeds unity in the most central events while, substantial deviations between the CMS measurements and several hydrodynamic model calculations are observed below 1\% centrality. The underlying cause of this deficit remains uncertain: it could arise from limitations in the modeling of the initial-state geometry or from an incomplete description of the medium’s hydrodynamic response.

Publication \cite{PhysRevC.102.054905} provides several explanations proposed as a remedy for the UCC puzzle. One possibility is that the {$^{208}$Pb} nuclei are not perfectly spherical but have intrinsic higher-order deformations, such as octupole shapes. In ultra-central collisions, these small deformations can dominate the geometry, enhancing the triangular flow. Another contribution comes from quantum fluctuations in nucleon positions, which create event-by-event geometric irregularities that are converted into flow through the hydrodynamic evolution of the quark–gluon plasma.

This article presents a systematic study of collective flow in ultra-central heavy-ion collisions, combining experimental measurements and theoretical interpretations to further investigate the properties of the quark--gluon plasma as a strongly interacting, nearly perfect fluid. The manuscript is organized as follows. Section 2 provides ALICE Run 3 detector upgrades and describes the event/track selection criteria employed in this work. In Section 3, the recent results are presented and discussed in comparison with theoretical model calculations. Finally, Section 4 summarizes the main conclusions and provides an outlook on future work.

\section{Experimental Details} 
During Long Shutdown 2 (2019–2022), the ALICE detector underwent a major upgrade to exploit the high-luminosity heavy-ion program of LHC Run 3 ~\cite{ALICE:Acharya_2024_upgrades}. The upgraded Inner Tracking System (ITS2) ~\cite{ALICE:2014ITS2TDR}, based on Monolithic Active Pixel Sensors (MAPS), provides significantly improved tracking and vertexing performance with a reduced material budget, while the Time Projection Chamber (TPC) ~\cite{ALICE:LHCC2013-020} was equipped with Gas Electron Multipliers (GEM) readout chambers enabling continuous operation ~\cite{ALICE:2013ReadoutTDR} at interaction rates up to 50 kHz. The Fast Interaction Trigger (FIT) ~\cite{ALICE:2017FIT} provides precise timing, triggering, and centrality determination. These detector upgrades, together with the new $O^2$ (Online-Offline) ~\cite{ALICE:2015O2TDR} computing framework for continuous readout, real-time calibration, and online reconstruction, allow ALICE to efficiently process the unprecedented data rates of Run 3.

Charged particles are reconstructed in the central barrel and selected within the kinematic range $0.2 < p_{\mathrm{T}} < 3.0$ GeV/$c$ and $|\eta| < 0.8$. The flow coefficients are measured using two- and multiparticle cumulant methods within the generic framework~\cite{Bilandzic:2014v2cumulants}. For the two-particle correlation, a pseudorapidity gap of $|\Delta\eta| > 1$ is applied to reduce non-flow contributions, such as correlations originating from jet fragmentation and resonance decays~\cite{Zhou:2015v34}. The ITS and the TPC detectors are used to reconstruct all charged particles for measuring $v_2$ and $v_3$. Since Run 3 operates with continuous TPC readout at very high interaction rates, multiple collisions and their associated particles are recorded simultaneously within the detector volume. This results in significantly increased detector occupancy, meaning that a large fraction of the readout channels are active at the same time. To reduce the impact of high-occupancy events on the reconstruction performance and data quality, an occupancy cut is applied. The relevant FIT detector component is FT0C. Minimum-bias events are selected by requiring at least one hit in both the FT0A and FT0C detector arrays within the nominal collision time window. Events are further rejected if the collision occurs too close to the boundaries of either the ITS readout frame or the time frame, ensuring high-quality event reconstruction. The event selection includes requirements on the position of the reconstructed primary $z$-vertex to ensure that the collision occurred within $\pm10$ cm of the nominal interaction point and within the fiducial acceptance of the central barrel detectors. Events affected by pile-up, corresponding to multiple collisions occurring within the same bunch crossing, are rejected using dedicated timing and vertex-based selection criteria.

The charged-particle tracks for the analysis are categorized into global tracks, ITS-only tracks, and TPC-only tracks. Global tracks are the best quality tracks that are matched between ITS and TPC. The TPC-only tracks are excluded. The selected track sample consists of global tracks satisfying standard track selection criteria (sel8), including requirements on the quality of the ITS and TPC track reconstruction, the number of associated detector clusters, the track-fit quality, and the distance of closest approach to the primary vertex. In addition, ITS-only tracks satisfying the corresponding ITS track selection criteria are included to improve the tracking efficiency. The combined sample of global and ITS-only tracks provides higher reconstruction efficiency, particularly at low transverse momentum. For the efficiency corrections, the sample of global plus ITS-only is constructed aiming higher efficiency. Residual non-uniformity in the TPC acceptance introduce distortions in the azimuthal distribution, leading to modulations that would not be present in an ideal detector with perfect cylindrical symmetry. To mitigate this effect, Non Uniform Acceptance (NUA) correction is applied at the track level. The detector efficiency is estimated using the general purpose Pb--Pb Monte Carlo production LHC24g3, anchored to the apass4 reconstruction conditions. More specifically, the Non Uniform Efficiency (NUE) is quantified as the ratio of the number of reconstructed tracks to the number of generated tracks providing a correction factor that can be applied to measured observables to account for detection and reconstruction losses.

\section{Results}

\begin{figure}[htbp!]
    \centering
    \begin{minipage}{0.49\textwidth}
        \centering
        \includegraphics[scale=0.32]{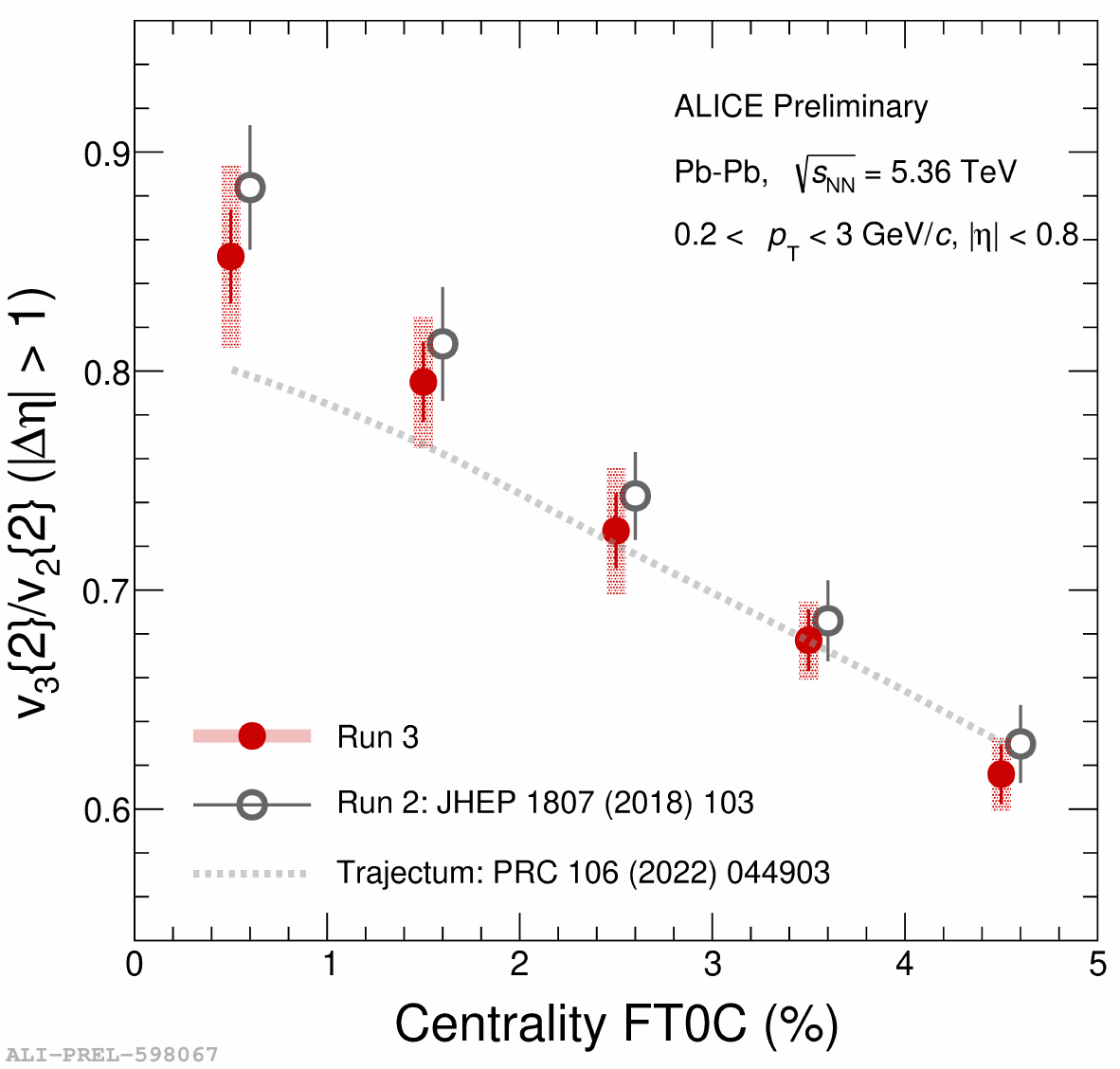}
        \caption{Triangular-to-elliptic flow ratio for central Pb--Pb collisions at $\sqrt{s_{\rm NN}}=5.36$ TeV. Red solid markers correspond to Run 3 data, hollow gray markers to Run 2 data, and the dashed gray band to the Trajectum model. Error boxes represent systematic uncertainties.}
        \label{fig:v32v22fulldata}
    \end{minipage}
    \hfill
    \begin{minipage}{0.49\textwidth}
        \centering
        \includegraphics[scale=0.32]{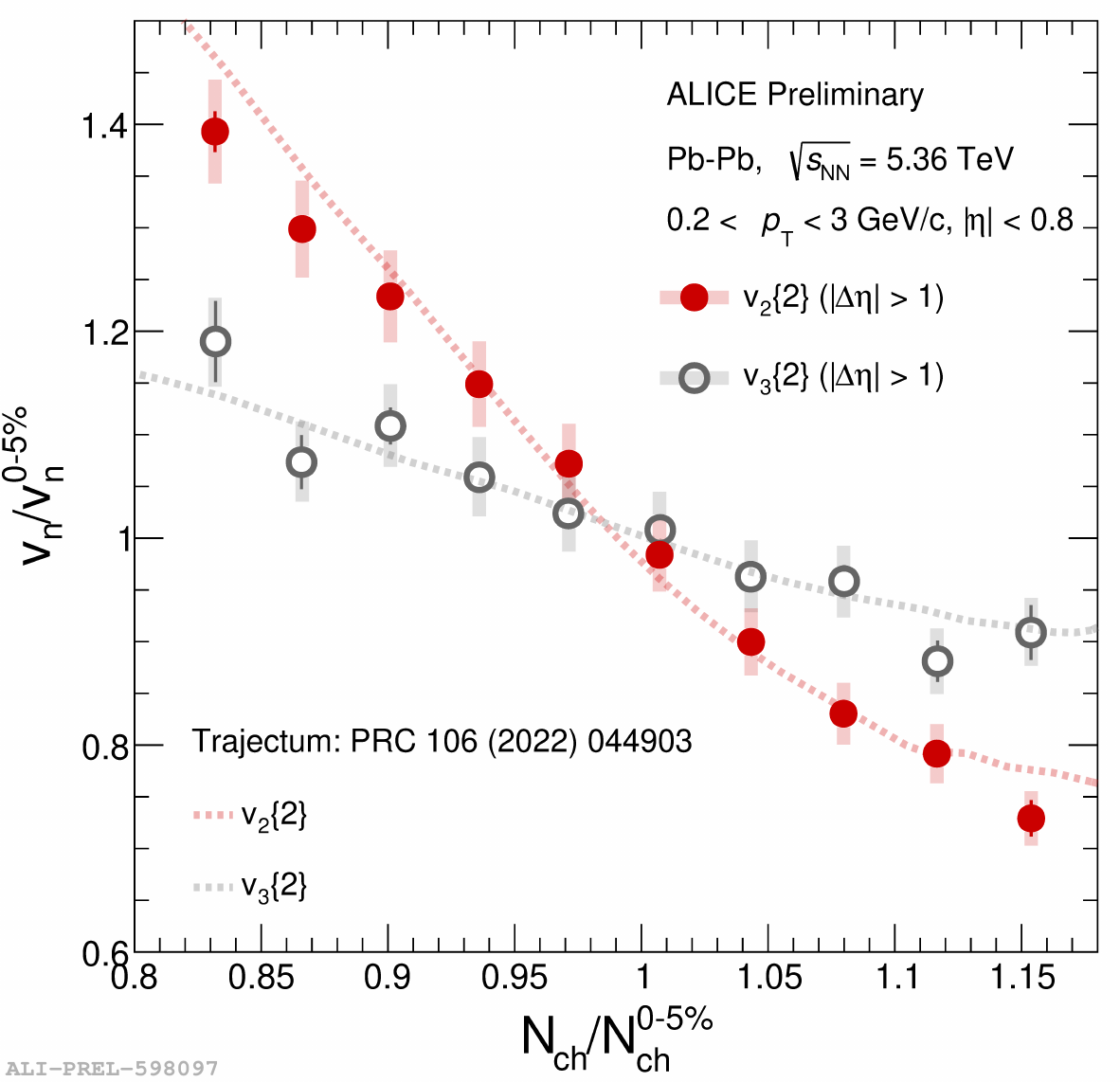}
        \caption{Self-normalized flow coefficients in central collisions. Red solid points correspond to $v_{2}\{2\}$ and hollow gray points to $v_{3}\{2\}$. The dashed red and gray lines show the corresponding Trajectum predictions.}
        \label{fig:vn_trajectum}
    \end{minipage}
\end{figure}

In this section, we present measurements of the ratio between triangular and elliptic flow in the 0--5\% centrality range. For comparison, the corresponding ALICE Run 2 results and predictions from the Trajectum hydrodynamic model are also shown. As illustrated in Fig.~\ref{fig:v32v22fulldata}, the Run 3 measurements are in good agreement with both the Run 2 results and the Trajectum calculations (up to 1\% centrality), demonstrating the consistency of the current analysis and the robustness of the measured flow observables.

Fig.~\ref{fig:vn_trajectum} presents the self-normalized elliptic and triangular flow coefficients as a function of the normalized charged-particle multiplicity. The measurements are compared with predictions from the Bayesian-tuned Trajectum model, which assumes a linear hydrodynamic response between the initial-state eccentricities and the final-state flow harmonics ($v_n \propto \varepsilon_n$). Overall, the model provides a good description of the measured trends. A particularly interesting feature is observed near the most central collisions. For $N_{\mathrm{ch}}/N_{\mathrm{ch}}^{0\text{--}5\%}<1$, the elliptic flow coefficient remains larger than the triangular flow coefficient, consistent with expectations from the average collision geometry. As the multiplicity approaches its maximum, however, the two harmonics intersect, and the triangular flow becomes larger than the elliptic flow. This behavior reflects the increasing importance of event-by-event fluctuations in ultra-central collisions, where the average elliptic geometry is strongly suppressed and fluctuation-driven triangular anisotropies become the dominant source of the observed collective flow.

\begin{figure}[htbp!]
    \centering
    \begin{minipage}{0.48\textwidth}
        \centering
        \includegraphics[scale=0.42]{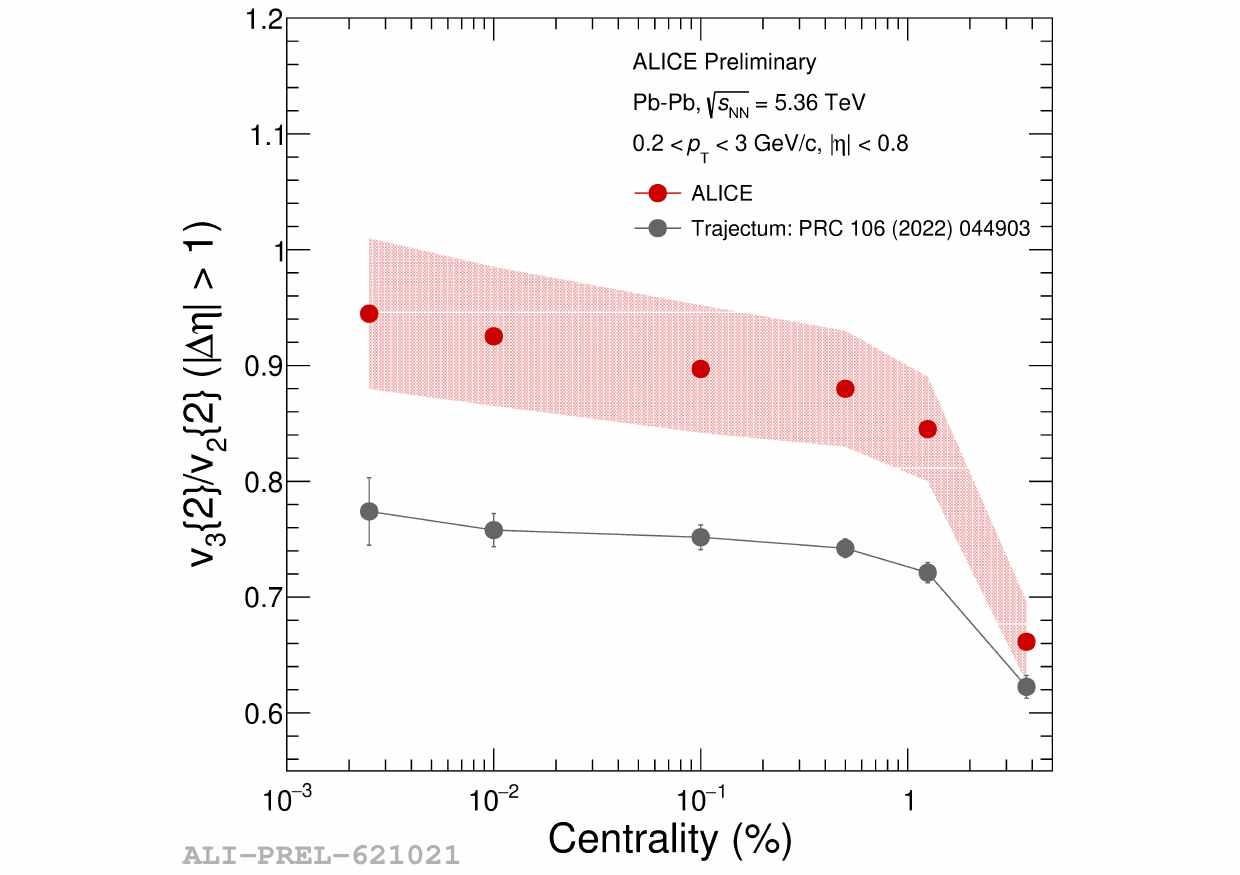}
        \caption{Flow ratio $v_{3}\{2\}/v_{2}\{2\}~(\left| \Delta\eta \right| > 1)$ in ultra-central Pb--Pb collisions. The centrality intervals are 0--0.005\%, 0--0.02\%, 0--0.2\%, 0--1\%, 0--2.5\%, and 2.5--5\%. The shaded band represents the systematic uncertainty.}
        \label{fig:ratiouccplot_v32v22}
    \end{minipage}
    \hfill
    \begin{minipage}{0.49\textwidth}
        \centering
        \includegraphics[scale=0.44]{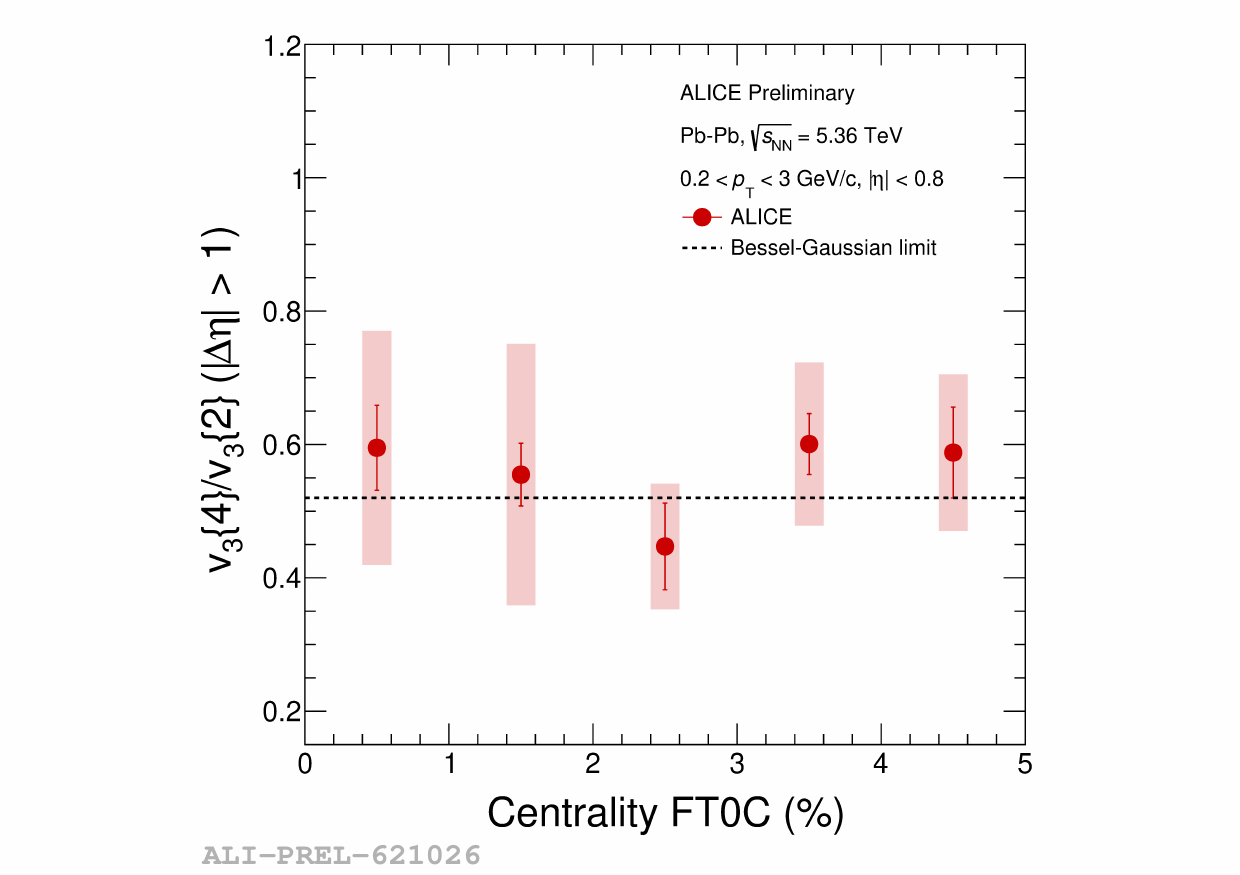}
        \caption{Ratio $v_{3}\{4\}/v_{3}\{2\}$ measured with $|\Delta\eta|>1$ in Pb--Pb collisions at 0--5\% centrality. The dashed line indicates the Bessel--Gaussian limit.} \vspace{2.8em}
        \label{fig:ratio05plotv34v32}
    \end{minipage}
\end{figure}

Probing towards ultra-central collisions, the corresponding ratio $v_{3}\{2\}/v_{2}\{2\}~(\left| \Delta\eta \right| > 1)$ is shown in Fig.~\ref{fig:ratiouccplot_v32v22}. The measurements obtained with ALICE Run 3 data are shown with statistical uncertainties represented by error bars (significantly smaller than the marker size) and systematic uncertainties indicated by shaded bands. Hydrodynamic predictions from the Trajectum model are included for comparison, providing a theoretical baseline for interpreting the observed behavior of the flow harmonics in ultra-central collisions. A clear discrepancy is observed between measured data and Trajectum.

To further investigate the origin of initial-state fluctuations, Fig.~\ref{fig:ratio05plotv34v32} presents the first ALICE Run 3 measurement of the ratio $v_{3}\{4\}/v_{3}\{2\}~(\left| \Delta\eta \right| > 1)$ in the 0--5\% centrality range. The measured ratio is consistent with the expectation from the Bessel--Gaussian (BG) limit, corresponding to $\sigma_{v_{3}}/\langle v_{3}\rangle \approx \sqrt{4/\pi-1}\approx0.52$. This result supports the interpretation that triangular flow in the 0--5\% centrality interval is predominantly driven by random initial-state fluctuations rather than by a coherent geometric contribution. 

\section{Summary} 

In summary, the measurements presented in this work highlight the evolving role of the initial-state geometry from central to ultra-central Pb--Pb collisions. In the 0--5\% centrality range, the collision geometry is still predominantly elliptic, reflecting the average overlap geometry of the two colliding nuclei. Consequently, the measured flow observables are well described by the Trajectum hydrodynamic model and are consistent with previous ALICE Run 2 measurements, as shown in Fig.~\ref{fig:v32v22fulldata}. The self-normalized flow measurements presented in Fig.~\ref{fig:vn_trajectum} further illustrate the increasing importance of event-by-event fluctuations as the most central collisions are approached. As the average elliptic geometry becomes progressively suppressed, the relative contribution of fluctuation-driven triangular flow increases, leading to the observed crossing of the normalized elliptic and triangular flow coefficients.

In the ultra-central regime, however, the measured $v_{3}\{2\}/v_{2}\{2\}$ ratio, shown in Fig.~\ref{fig:ratiouccplot_v32v22}, is no longer reproduced by the Trajectum model. This discrepancy suggests that the current description of the initial state is incomplete and that higher-order geometric fluctuations, or an additional coherent source of triangular anisotropy, may contribute to the observed collective flow. One possible explanation is the presence of an intrinsic octupole deformation of the $^{208}$Pb nucleus.

Lastly, the first ALICE Run 3 measurement of the ratio $v_{3}\{4\}/v_{3}\{2\}$ in the 0--5\% centrality range was presented. As shown in Fig.~\ref{fig:ratio05plotv34v32}, the measured ratio is consistent with the Bessel--Gaussian limit, indicating that triangular flow in this centrality interval is predominantly driven by event-by-event fluctuations, with no evidence for geometry-driven contribution. This result establishes an important baseline for future measurements in increasingly central collisions. Future work will extend these measurements to the ultra-central region and compare them with hydrodynamic calculations incorporating fluctuating octupole deformation of the $^{208}$Pb nucleus. Such studies may provide a unified description of both the enhancement of the $v_{3}\{2\}/v_{2}\{2\}$ ratio and the evolution of $v_{3}\{4\}/v_{3}\{2\}$, offering new insight into the interplay between initial-state fluctuations, intrinsic nuclear structure, and the hydrodynamic response of the quark--gluon plasma.

\bibliographystyle{elsarticle-num}
\bibliography{sqm2026_template}

@article{Ollitrault1992, 
title = {Anisotropy as a signature of transverse collective flow}, 
author = {Ollitrault, Jean-Yves}, 
journal = {Phys. Rev. D}, 
volume = {46}, 
issue = {1}, 
pages = {229--245},
numpages = {0}, 
year = {1992}, 
month = {7}, 
publisher = {American Physical Society},
doi = {10.1103/PhysRevD.46.229}
}

@article{Heinz2013FlowReview,
  author = {Heinz, Ulrich and Snellings, Raimond},
  title = {Collective flow and viscosity in relativistic heavy-ion collisions},
  journal = {Annual Review of Nuclear and Particle Science},
  volume = {63},
  pages = {123--151},
  year = {2013},
  eprint = {1301.2826}
}

@article{Voloshin:1994mz,
  author        = {Voloshin, S. and Zhang, Y.},
  title         = {Flow Study in Relativistic Nuclear Collisions by Fourier Expansion of Azimuthal Particle Distributions},
  journal       = {Z. Phys. C},
  volume        = {70},
  pages         = {665--672},
  year          = {1996},
  eprint        = {hep-ph/9407282},
  archivePrefix = {arXiv},
  primaryClass  = {hep-ph}
}

@article{PhysRevC.102.054905,
  author        = {Carzon, P. and Rao, S. and Luzum, M. and Sievert, M. and Noronha-Hostler, J.},
  title         = {Possible octupole deformation of $^{208}\mathrm{Pb}$ and the ultracentral $v_{2}$ to $v_{3}$ puzzle},
  journal       = {Phys. Rev. C},
  volume        = {102},
  number        = {5},
  pages         = {054905},
  year          = {2020},
  month         = nov,
  eprint        = {2007.06586},
  archivePrefix = {arXiv},
  primaryClass  = {nucl-th}
}

@article{ALICE:Adam_2016,
  author        = {{ALICE Collaboration}},
  collaboration = {ALICE},
  title         = {Anisotropic Flow of Charged Particles in $\mathrm{Pb}$--$\mathrm{Pb}$ Collisions at $\sqrt{s_{\mathrm{NN}}}=5.02$ $\mathrm{TeV}$},
  journal       = {Phys. Rev. Lett.},
  volume        = {116},
  number        = {13},
  pages         = {132302},
  year          = {2016},
  month         = apr,
  eprint        = {1602.01119},
  archivePrefix = {arXiv},
  primaryClass  = {nucl-ex}
}

@article{Giannini_2023,
   author        = {Giannini, A. V. and Ferreira, M. N. and Hippert, M. and Chinellato, D. D. and Denicol, G. S. and Luzum, M. and Noronha, J. and Nunes da Silva, T. and Takahashi, J.},
  title         = {Assessing the ultracentral flow puzzle in hydrodynamic modeling of heavy-ion collisions},
  journal       = {Phys. Rev. C},
  volume        = {107},
  number        = {4},
  pages         = {044907},
  year          = {2023},
  month         = apr,
  eprint        = {2209.14001},
  archivePrefix = {arXiv},
  primaryClass  = {nucl-th}
}

@article{ALICE:Acharya_2024_upgrades,
  author        = {{ALICE Collaboration}},
  collaboration = {ALICE},
  title         = {$\mathrm{ALICE}$ upgrades during the $\mathrm{LHC}$ Long Shutdown 2},
  journal       = {JINST},
  volume        = {19},
  number        = {05},
  pages         = {P05062},
  year          = {2024},
  month         = may,
  eprint        = {2401.13727},
  archivePrefix = {arXiv},
  primaryClass  = {physics.ins-det}
}

@article{ALICE:2014ITS2TDR,
  author        = {{ALICE Collaboration}},
  collaboration = {ALICE},
  title         = {Technical Design Report for the Upgrade of the $\mathrm{ALICE}$ Inner Tracking System},
  journal       = {J. Phys. G},
  volume        = {41},
  number        = {8},
  pages         = {087002},
  year          = {2014},
  eprint        = {1402.4476},
  archivePrefix = {arXiv},
  primaryClass  = {physics.ins-det}
}

@techreport{ALICE:LHCC2013-020,
  author        = {{ALICE Collaboration}},
  collaboration = {ALICE},
  title         = {Technical Design Report for the Upgrade of the $\mathrm{ALICE}$ Time Projection Chamber},
  institution   = {CERN},
  address       = {Geneva},
  number        = {CERN-LHCC-2013-020, ALICE-TDR-016},
  year          = {2014},
  month         = jan,
  type          = {Technical Design Report},
  url           = {https://cds.cern.ch/record/1622286}
}

@article{ALICE:2017FIT,
  author        = {Trzaska, W. H. and others},
  collaboration = {ALICE},
  title         = {New Fast Interaction Trigger for $\mathrm{ALICE}$},
  journal       = {Nucl. Instrum. Meth. A},
  volume        = {845},
  pages         = {463--466},
  year          = {2017},
  doi           = {10.1016/j.nima.2016.05.016}
}

@techreport{ALICE:2015O2TDR,
  author        = {{ALICE Collaboration}},
  collaboration = {ALICE},
  title         = {Technical Design Report for the Upgrade of the Online--Offline Computing System},
  institution   = {CERN},
  address       = {Geneva},
  number        = {CERN-LHCC-2015-006, ALICE-TDR-019},
  type          = {Technical Design Report},
  year          = {2015},
  month         = mar,
  url           = {https://cds.cern.ch/record/2011297}
}

@techreport{ALICE:2013ReadoutTDR,
  author        = {{ALICE Collaboration}},
  collaboration = {ALICE},
  title         = {Upgrade of the $\mathrm{ALICE}$ Readout and Trigger System},
  institution   = {CERN},
  address       = {Geneva},
  number        = {CERN-LHCC-2013-019, ALICE-TDR-015},
  type          = {Technical Design Report},
  year          = {2013},
  month         = sep,
  url           = {https://cds.cern.ch/record/1603472}
}

@article{Bilandzic:2014v2cumulants,
  author        = {Bilandzic, A. and Christensen, C. H. and Gulbrandsen, K. and Hansen, A. and Zhou, Y.},
  title         = {Generic Framework for Anisotropic Flow Analyses with Multiparticle Azimuthal Correlations},
  journal       = {Phys. Rev. C},
  volume        = {89},
  number        = {6},
  pages         = {064904},
  year          = {2014},
  month         = jun,
  eprint        = {1312.3572},
  archivePrefix = {arXiv},
  primaryClass  = {nucl-ex}
}

@article{Zhou:2015v34,
  author        = {Zhou, Y. and Zhu, X. and Li, P. and Song, H.},
  title         = {Investigation of possible new features of anisotropic flow in relativistic heavy-ion collisions},
  journal       = {Phys. Rev. C},
  volume        = {91},
  number        = {6},
  pages         = {064908},
  year          = {2015},
  month         = jun,
  eprint        = {1501.06992},
  archivePrefix = {arXiv},
  primaryClass  = {nucl-th}
}



\end{document}